\documentclass[
 reprint,
 superscriptaddress,
 amsmath,amssymb,
 aps,
pra,
showkeys
]{revtex4-2}

\usepackage{soul}
\usepackage{float}
\usepackage{multibib}
\usepackage{hyperref}
\usepackage{xcolor}
\hypersetup{
    colorlinks,
    linkcolor={blue!50!blue},
    citecolor={blue!50!blue},
    urlcolor={blue!80!black}
}
\usepackage{siunitx}
\usepackage{import}
\usepackage{placeins}
\usepackage{xcolor}
\usepackage{lmodern}
\usepackage[braket, qm]{qcircuit}
\usepackage{graphicx}
\usepackage{dcolumn}
\usepackage{bm}
\usepackage{comment}
\usepackage[english]{babel}
\usepackage{blindtext}
\usepackage{physics}  
\usepackage{multirow}
\usepackage{array}
\usepackage{enumerate}

\usepackage{ulem}
\definecolor{myblue}{rgb}{0.0, 0.75, 1.0}
\definecolor{lightpink}{rgb}{0.9, 0.4, 0.38}

\begin{document}
\title{Identifying electronic transitions of defects in hexagonal boron nitride for quantum memories}

\author{Chanaprom Cholsuk}
\email{chanaprom.cholsuk@tum.de}
\affiliation{Department of Computer Engineering, School of Computation, Information and Technology, Technical University of Munich, 80333 Munich, Germany}
\affiliation{Abbe Center of Photonics, Institute of Applied Physics, Friedrich Schiller University Jena, 07745 Jena, Germany}

\author{Asl{\i} \surname{\c{C}akan}}
\affiliation{Department of Computer Engineering, School of Computation, Information and Technology, Technical University of Munich, 80333 Munich, Germany}

\author{Sujin Suwanna}
\affiliation{Optical and Quantum Physics Laboratory, Department of Physics, Faculty of Science, Mahidol University, Bangkok 10400, Thailand}.

\author{Tobias Vogl}%
\email{tobias.vogl@tum.de}
\affiliation{Department of Computer Engineering, School of Computation, Information and Technology, Technical University of Munich, 80333 Munich, Germany}
\affiliation{Abbe Center of Photonics, Institute of Applied Physics, Friedrich Schiller University Jena, 07745 Jena, Germany}

\date{\today}

\begin{abstract}
A quantum memory is a crucial keystone for enabling large-scale quantum networks. Applicable to the practical implementation, specific properties, i.e., long storage time, selective efficient coupling with other systems, and a high memory efficiency are desirable. Though many quantum memory systems are developed thus far, none of them can perfectly meet all requirements. This work herein proposes a quantum memory based on color centers in hexagonal boron nitride (hBN), where its performance is evaluated based on a simple theoretical model of suitable defects in a cavity. Employing density functional theory calculations, 257 triplet and 211 singlet spin electronic transitions are investigated. Among these defects, it is found that some defects inherit the $\Lambda$ electronic structures desirable for a Raman-type quantum memory and optical transitions can couple with other quantum systems. Further, the required quality factor and bandwidth are examined for each defect to achieve a 95\% writing efficiency. Both parameters are influenced by the radiative transition rate in the defect state. In addition, inheriting triplet-singlet spin multiplicity indicates the possibility of being a quantum sensing, in particular, optically detected magnetic resonance. This work therefore demonstrates the potential usage of hBN defects as a quantum memory in future quantum networks.
\end{abstract}


\maketitle

\section{Introduction}
\begin{figure*}[ht]
    \centering
    \includegraphics[width = 0.8\textwidth]{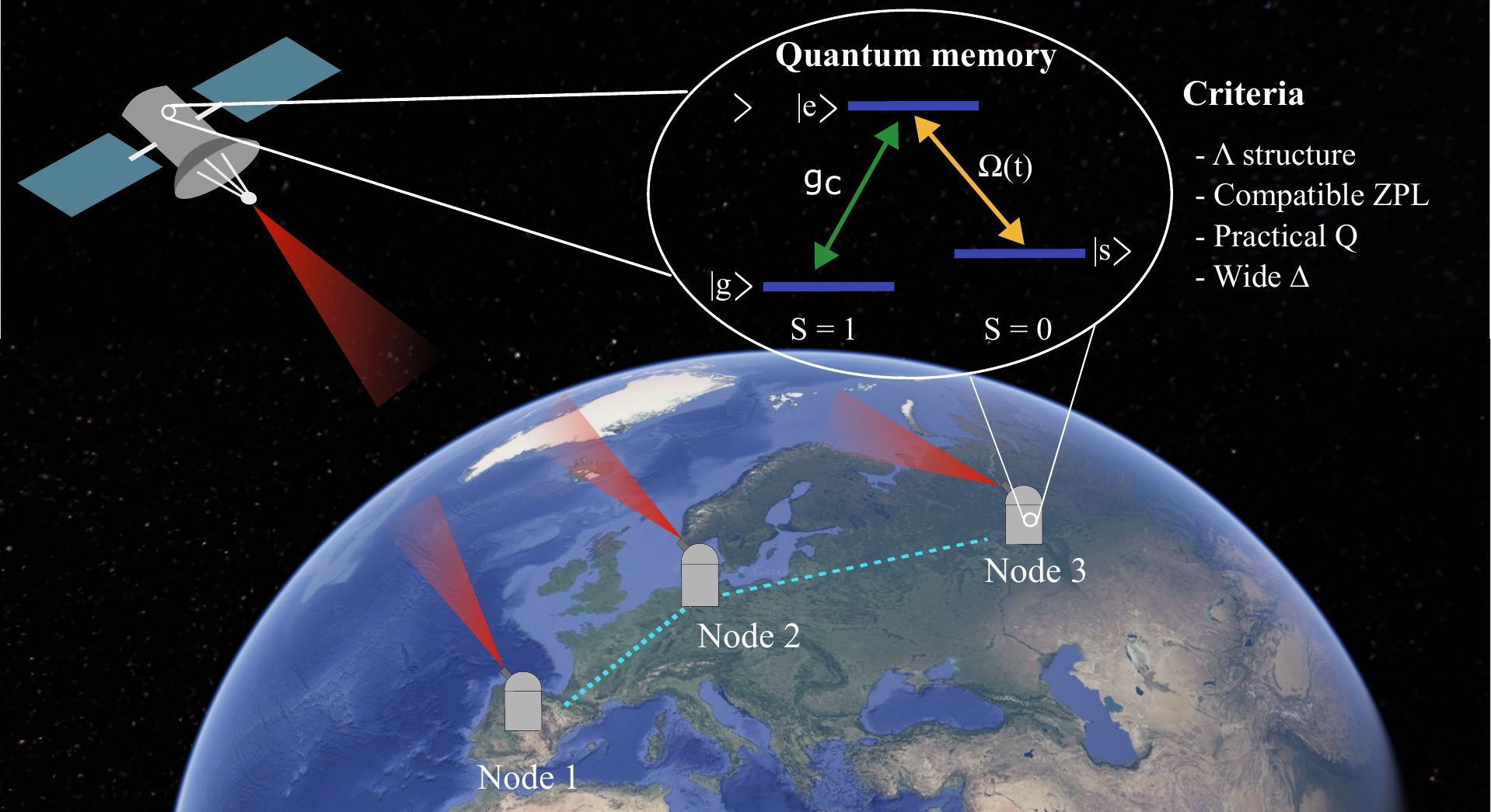}
    \caption{Schematics of quantum networks distributed by quantum repeaters in which each node consists of the quantum memory as one of the components where the $\Lambda$-structure quantum memory needs to satisfy our four established criteria. In a global quantum network, there can be quantum memories on satellites or on the ground \cite{10.1038/s41534-021-00460-9}, likely in a hybrid way with a combination of both.}
    \label{fig:overview}
\end{figure*}

A quantum memory is an essential building block enabling quantum repeaters and thus potentially revolutionizing large-scale quantum technologies, comprising of quantum communication \cite{10.1038/s41534-021-00460-9,Muralidharan2016,quick3,Mostafa-qkd}, networked quantum computing \cite{Cuomo2020,Gyongyosi2021}, and quantum-enhanced sensing \cite{Gilmore2021}. Its successful implementation among each building block is to communicate among them over long distances; for instance, based on entanglement swapping via the Bell state measurement (BSM), which requires flying qubits from photons \cite{PhysRevLett.70.1895,PhysRevLett.80.1121,Duan2001,gottesman2012longer,Lee2013BellstateMA,PhysRevLett.81.5932}. Consequently, a quantum memory plays an essential role in storing quantum states encoded in incoming photons,  transforming them into matter qubits, holding for an amount of time, and later converting them on demand back to the photonic states. Most recently, the growing interest in feasible quantum memory provides instances for warm vapour memories \cite{Kaczmarek2018}, laser-cooled atomic systems \cite{Cho2016}, rare-earth ion doped crystals \cite{Rivera}, and solid-state crystals \cite{Bradley2022,SiCmemory,Udvarhelyi2022}. Common challenges of these physical systems include the capability to store a quantum state, coherently control and retrieve, and efficiently couple to other quantum systems.\\
\indent Solid-state quantum memories, particularly for nitrogen-vacancy centers in diamond \cite{Childress2013, Bradley2022, Pfender2017, Fuchs2011, Ji2022}, transition metal defects in silicon carbide \cite{SiCmemory}, and the carbon center in silicon \cite{Udvarhelyi2022} demonstrate the potential capabilities to tackle these challenges. These systems require two key properties: the $\Lambda$-shaped electronic structure and the storing state. The $\Lambda$ structure is determined by the optical transitions of the defect system. The aforementioned solid-state crystals feature the intersystem-crossing transition between the triplet and singlet configurations. Purposefully some nuclear spins can be used for storage due to significantly longer coherent time than that of the electron spin. Nevertheless, these systems still have some disadvantages, such as undesired phonon transition in the NV center from thermal noise \cite{Reiserer2016} or even the requirement of pure quantum states for the ground-state hyperfine manifold in SiC \cite{SiCmemory}. Although recent progress on these systems has been significant and extensive, for instance, detuning the control laser far from the phonon side band (PSB) \cite{Reiserer2016}, or even proposing a novel nuclear spin preparation protocol for SiC \cite{SiCmemory}, further development still necessitates and remains a study of active research.\\
\indent In this work, we investigate feasible candidates for a quantum memory in fluorescent defects localized in a two-dimensional hexagonal boron nitride (hBN). We explored hBN defects because of the unique promising properties of hBN that could potentially endow advantages for a quantum memory; namely, it has low Fresnel reflection \cite{Tran2016}, relatively low phonon coupling \cite{10.1021/acsphotonics.8b00127}, multiple possible defect choices for the $\Lambda$ structure, coupling with the nuclear spin, and integrating with other quantum-network components \cite{Cholsuk2022, cholsuk2023comprehensive}. This may potentially alleviate the ongoing issues occurring in the bulk systems \cite{Childress2013, Bradley2022, Pfender2017, Fuchs2011, Ji2022,SiCmemory,Udvarhelyi2022}. Since the consideration of hBN is lacking pertaining to alternative applications of quantum memory so far, this work paves a way forward towards the aims to (1) investigate the feasibility and performance of hBN defects for being a quantum memory by calculating the writing efficiency, quality factor, bandwidth, and lifetime based on a simple dynamical quantum memory model, and (2) classify electronic transitions of 257 triplet and 211 singlet configurations and collect them into a database for further experiments.\\
\indent There have been several protocols for quantum memories, such as electromagnetically induced transparency (EIT) \cite{Heinze2013}, Raman off-resonance scattering \cite{Zheltikov2018}, and controlled reversible inhomogeneous broadening (CRIB) \cite{PhysRevA.83.012318}. This work assumes the tantamount mechanism as the Raman (off-resonance) quantum scattering protocol. More precisely, the hBN defects must inherit the $\Lambda$ structure to have the transition between triplet and singlet spin configurations via an inter-system crossing channel. Then, the signal and control pulses are expected to couple with a photon and function during the writing/reading process. In some mechanisms, like in the bulk systems \cite{Childress2013, Bradley2022, Pfender2017, Fuchs2011, Ji2022,SiCmemory,Udvarhelyi2022}, the information will later be stored in the nuclear spin by entangling with the electron spin. Nonetheless, the existence of the $\Lambda$ structure is still indispensable. Due to the lack of identification of hBN for the application of quantum memories, our model was applied to first evaluate the performance of hBN being the $\Lambda$ structure, a fundamental prerequisite for quantum memory. This includes ground state ($\ket{g}$), excited state ($\ket{e}$), and meta-stable state ($\ket{s}$). Here, we assigned the meta-stable state $\ket{s}$ as a storage state to estimate the writing efficiency. This way, we aim to characterize a large number of defects and rule some of them out based on our construction, which consists of four criteria: inheriting the $\Lambda$ structure, compatible ZPL with other coupling systems, required cavity quality factor in practice, and wide bandwidth as depicted in Fig.~\ref{fig:overview} and discussed in detail in Sec.~\ref{sec:criteria}. With the selection of our model, these defects are already qualified to be later applied to any other quantum memory mechanism, such as coupling with the nuclear spin. However, we leave the research questions of which the optimal nuclear spin is and how we couple nuclear and electron spins in hBN for our ongoing work.\\
\indent As our database reveals the intrinsic triplet-singlet defects and includes transition energy, zero-phonon line (ZPL), transition dipole moment ($\mu$), lifetime $(\tau)$, quality factor ($Q$), and bandwidth ($\Delta$). This is not only useful for quantum memory applications, but also applies to quantum sensing applications, including optically detected magnetic resonance (ODMR) with the ground- \cite{Stern2022,Zhou2023} and excited states \cite{Mu2022}. The ODMR effect can be even used to sense strain, as strain shifts the energy levels \cite{Lyu2022}. Using the ODMR technique is potentially more sensitive than estimating the strain based on the shift of the ZPL. Even though there are some 2D databases reported earlier \cite{Bertoldo2022, Sajid2023}, the hBN defects in particular remain largely unexplored and were calculated based on the common Perdew-Burke-Ernzerhof (PBE) functional, which might underestimate the bandgap and transition energy, especially for the wide-band gap material like hBN \cite{Reimers2018,Sajid2018-EPR}. Our database on the other hand is then filling this critical discrepancy and also provides a new dataset calculated from a more accurate HSE06 functional than PBE. This work therefore contributes to the realization of hBN as a quantum memory through our material-free quantum memory model and also provides a comprehensive database, which can be made use of by other quantum applications.

\section{Methodology}
In this section, we first summarize important parameters used for constructing the quantum memory model, following our initial work in Ref.~\cite{Takla2023}. Then, the details of the first-principle calculation and database collection are explained. Last, the criteria for determining performance are elaborated.
\subsection{Quantum memory model}
\subsubsection{Electronic transition in quantum memory}
The performance of the proposed quantum memory model is simulated based on the assumptions that the electronic transition takes place among the $\Lambda$ three-level structures: a ground state $\ket{g}$, an excited state $\ket{e}$, and a meta-stable state $\ket{s}$. Here, we summarized important parameters implemented in this work. The Hamiltonian governing such transition is given by
\begin{equation}
    H = \hbar\pqty{\Omega\hat{\sigma}_\text{se}\otimes\hat{1} + g\hat{\sigma}_\text{ge}\otimes \hat{a} +\text{H.c.}} + \Delta\hat{\sigma}_\text{ee}\otimes\hat{1} + \delta\hat{\sigma}_\text{ss}\otimes\hat{1},
    \label{eq:Hamiltonian}
\end{equation}
where an incoming photon between $\ket{g}$ and $\ket{e}$ is coupled by a signal field with the coupling constant $g_c$. A control pulse between $\ket{e}$ and $\ket{s}$ is restricted into a sigmoid function with characteristic time $T$ as shown in Eq.~\ref{eq:control-pulse} for an example. Of course, other functions can also be applied. 
\begin{equation}
    \Omega(t) = \frac{\Omega_0}{1+\exp\frac{t}{T}},
    \label{eq:control-pulse}
\end{equation}
where $\Omega_0$ is a peaked Rabi frequency. The transition is expressed by the atomic operator $\hat{\sigma}_{ij} = \ket{i}\bra{j}$. $\Delta$ and $\delta$ represent the one-photon detuning from the $\ket{g}$ to $\ket{e}$ transition, and a two-photon detuning from the $\ket{g}$ to $\ket{s}$ transition, respectively. In our current model, we neglect both detunings.\\
\indent To map between the initial atomic state and a photon, it is done via the stimulated Raman adiabatic passage (STIRAP) process, which utilizes the changes of the dark state $\Phi_0$ to change the wavefunction $\ket{\Psi}$ via the mixing angle $\theta$ \cite{Vitanov2017,Koerber2017}. Under the adiabatic process, $\ket{\Psi}$ needs to change sufficiently slow in which it maintains in the dark state. The dark state $\ket{\Phi_0}$ can then be expressed as
\begin{align}
    \ket{\Phi_0} &= \cos \theta \ket{g,1} + \sin \theta \ket{s,0}\\
    \theta\ &= \arctan \frac{g}{\Omega(t)},
\end{align}
where $\ket{\Phi_0}$ is written in the form of composite system $\ket{\Psi} = \ket{\text{atomic state}}\otimes\ket{\text{photon number}}$. \\
\indent Further, we assume that the quantum memory is in a cavity with a volume $V$; thus, the coupling constant ($g_c$) between an incoming photon and a signal field can be computed by \cite{Gorshkov2007-1}
\begin{equation}
    g_c = \bra{e}\hat{d}_{ge}\cdot \epsilon \ket{g} \sqrt{\frac{\omega_1}{2\hbar \epsilon_0 V}},
    \label{eq:couplingConstant-g}
\end{equation}
where $V$ is the volume of the cavity, which scales as $V = n\lambda^3$ where $n = 1.76$ is used in our calculation and it was taken from an experiment of a hBN defect coupled to a microcavity \cite{Vogl2019}; $\hat{d}_{ge}$ is a transition dipole moment obtained from density functional theory (DFT) calculation from $\ket{g}$ to $\ket{e}$; and $\epsilon$ is a polarization unit vector of the signal field.
\subsubsection{Quantum memory performance}
The dynamics of a quantum memory are simulated by the Lindblad master equation, which takes population decay rates into account and is given by
\begin{equation}
    \frac{d}{dt}\rho = -i[H,\rho] + \sum_{i,j} C_{ij}\rho C^\dagger_{ij} - \frac{1}{2} \anticommutator{C^\dagger_{ij} C_{ij}}{\rho},
    \label{eq:masterEq_decay}
\end{equation}
where $C_{ij} = \sqrt{\gamma_{ij}}\ketbra{i}{j}$ denotes a jump operator between states $i$ and $j$. In our case, these states are ground, excited, and meta-stable states: $i,j \in \{g, e, s\}$. $\gamma_{ij}$ denotes the decay rate between these channels. As the $\ket{s}$ state plays a critical role as a storage, we calculated the writing efficiency by computing the probability of the transition from $\ket{g}$ to $\ket{s}$ using the QuTiP package. It is noted that the $\ket{s}$ is selected as a storing state because the transition between $\ket{s}$ and $\ket{g}$ is unlikely. Also, in the absence of any decays, the writing efficiency will be equal to the reading efficiency because of the reversibility of the transition pathways. Despite this reversible transition pathway, the transition between $\ket{s}$ and $\ket{g}$ can still exist in principle (even though they are dipole-forbidden) via another intersystem crossing. Whether this is relevant depends on the required storage time, i.e., the lifetime of this transition needs to be longer than the storage time; otherwise the stored quantum information is lost through relaxation into the ground state.

\subsubsection{Mapping materials properties into quantum memory performance}
To evaluate the quantum memory performance based on the material's properties, we computed the quality factor of a cavity and the bandwidth corresponding to each defect, which can achieve 95\% writing efficiency. To that end, we first calculated the dark-state population $d(t)$ among $\ket{g}$, $\ket{e}$, and $\ket{s}$ affected by the decay rate of the cavity with the following equation. 
\begin{equation}
    \kappa(t) = k \cos^2\theta(t),
\end{equation}
where $k$ is the non-zero value causing a loss of photons in the cavity.
The governing equation for the dark-state population decay under the adiabatic process becomes
\begin{align}
    \dv{}{t} d(t) &= -\kappa(t) d(t) = -k \cos^2\theta(t) d(t). \label{eq:ode1}
\end{align}
According to the control field $\Omega (t)$ given by Eq.~\eqref{eq:control-pulse}, the mixing angle $\theta$ can then be substituted and given as
\begin{eqnarray}
    \theta = \arctan\left({\frac{g_c}{\omega_0}\left(1+\exp{\frac{t}{T}}\right)}\right).
\end{eqnarray}
Eq.~\eqref{eq:ode1} can be solved analytically and leads to
\begin{eqnarray}
        d(t) =c_1 &\exp& \pqty{\frac{k\Omega _0^2 \left(T \log \left(\left(e^{t/T}+1\right)^2+\Omega _0^2\right)-2 t\right)
        }{2 \left(\Omega _0^2+1\right)}} \nonumber \\
        \times &\exp& \left(\frac{2k  \Omega _0 T \tan ^{-1}\left(\frac{e^{t/T}+1}{\Omega _0}\right)}{2 \left(\Omega _0^2+1\right)}\right),
\end{eqnarray}
where $c_1$ is a constant, depending on the initial condition. Here, we set the initial dark-state population $d(t_0) = d_0 = 0.999$ at a particular $t_0$, implying that a photon enters the cavity when $99.9\%$ of the dark-state population is in the state $\ket{g}$. This condition is trivially fulfilled for a suitable quantum emitter system at room temperature (i.e., the Fermi energy is not close to the excited state). We note that the initial time $t_0$ can be obtained by
\begin{equation}
    t_0 = T \log \pqty{\frac{p \abs{\Omega_0} }{\sqrt{p(1-p)}}-1},
    \label{eq:time_in_sigmoid}
\end{equation}
where $p$ is the probability of the dark state overlapping with the metastable state. Finally, we need to find the decay rate that satisfies the condition of each writing efficiency.\\
\indent Once we obtain the satisfied decay rate, the quality factor $Q$ required for each particular defect can be computed by
\begin{equation}
    Q = \frac{\omega}{2 \kappa},
    \label{eq:decay-quality}
\end{equation}
where $\omega$ is the resonance frequency between the $\ket{g}$ and $\ket{e}$ transition, corresponding to the ZPL.\\
\indent Finally, the bandwidth $\Delta$ for each defect is also calculated to capture the working range of our memory under a desired threshold of writing efficiency. For this, we calculate the half-width at half maximum ($\sigma_\Delta$) of the writing efficiency affected by one-photon detuning, which yields
\begin{equation}
    \Delta = g_c \sigma_\Delta.
    \label{eq:simplify-bandwidth}
\end{equation}
See Ref.~\cite{Takla2023} for additional details of the derivation.
\begin{figure*}[ht]
    \centering
    \includegraphics[width = 1\textwidth]{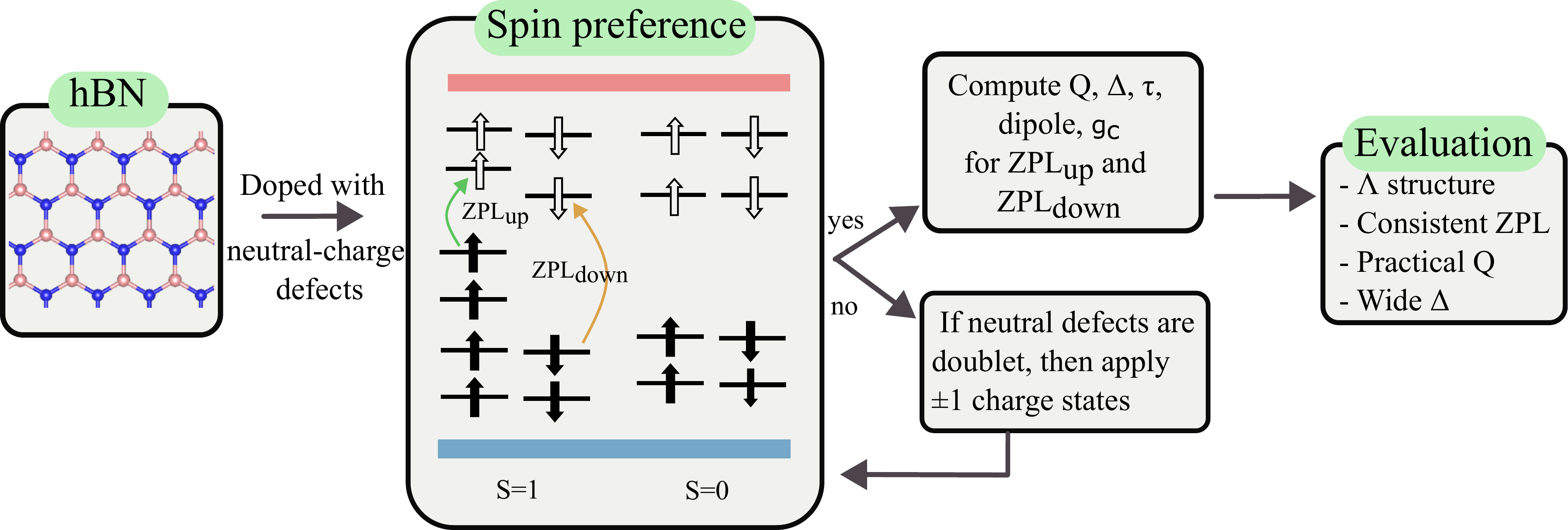}
    \caption{Flowchart of the DFT database collection. The filled (unfilled) up and down arrows indicate the occupied (unoccupied) defect states with spin-up and spin-down polarization, respectively. The parameters $Q, \Delta, \tau,$ and $g_c$ indicate the quality factor, bandwidth, lifetime, and coupling constant, respectively.}
    \label{fig:workflow}
\end{figure*}
\subsection{First-principle calculation}
\subsubsection{Simulating triplet-singlet spin multiplicity}
\indent All DFT calculations with spin polarization included in this work are performed by the Vienna Ab initio Simulation Package (VASP) \cite{vasp1,vasp2}. A plane wave is used as a basis set to handle the periodic structures of 7$\times$7$\times$1 supercell size. Pseudopotentials treated by the projector augmented wave (PAW) method are chosen to account for the nucleus and valence electrons \cite{paw,paw2}. Additional computational details can be found in Supplementary Section S1.\\
\indent As it is well known that hBN has a large band gap ($\sim$6 eV) and the conventional PBE functional typically underestimates it, we circumvent this issue by employing HSE06 functional instead, as it has been benchmarked well with experimental results \cite{Reimers2018,Sajid2018-EPR}. We remark that every DFT strategy always has finite accuracy due to defect environment and interaction; for example, the HSE06 functional has an uncertainty by $\sim$0.3-0.8 eV, depending on defect electronic states \cite{Reimers2020}. Intrinsic crystal strain can shift the properties, especially for the defects fabricated by ion implantation due to some substantial residual strain. Thus, for the purpose of the microscopic assignment to identify the responsible defect or its photophysical properties, comparing results across different DFT schemes, i.e., periodic model and molecular cluster model, or comparing with various properties \cite{Reimers2020,cholsuk2023comprehensive} may be worthwhile. However, as this work investigates 468 electronic structures in total, comparing results with several DFT techniques requires huge computational demand. We thus compared our calculation with other DFT calculations and our own experiments \cite{Abdi2018,Jara2020,Chen2021,Sajid2018,cholsuk2023comprehensive,anand-dipole} as demonstrated in Supplementary Section S3. As such, we have restricted ourselves to employing only HSE06 functional but still provided the fingerprint of all defects for benchmarking with the future experiment.\\
\indent To consider spin multiplicity, we relaxed every structure and simultaneously constrained the total spin of all defects using NUPDOWN = 0, 1, and 2 in VASP for singlet (S = 0), doublet (S = 1/2), and triplet (S = 1) spin configurations. If the electron is fully occupied on both singlet and triplet configurations, that defect satisfies our defect screening criteria. Then, such defects are investigated further in the excited-state configurations by manually occupying electrons in the excited state based on the $\Delta$SCF method \cite{Jones1989}. However, if the neutral-charge defect has a full electron occupation in the doublet configuration instead, singly positive and negative charges will be applied, as summarized in Fig.~\ref{fig:workflow}. As the optical transition requires two-level defect states with conserved spin polarization and in principle can take place in both spin-up and spin-down pathways, we then investigated the ZPL, transition dipole moment ($\boldsymbol{\mu}$), and lifetime of both spin up and spin-down transitions if the spin-conserved two-level defect states exist.

\subsubsection{Transition dipole moment}
The transition dipole moment ($\boldsymbol{\mu}$) is calculated from the $\ket{g}$ to $\ket{e}$ states to represent the coupling constant of the signal field $g_c$ by
\begin{equation}
\boldsymbol{\mu} = \frac{i\hbar}{(E_{f} - E_{i})m}\bra{\psi_{f}}\textbf{p}\ket{\psi_{i}},
\label{eq:dipole}
\end{equation}
where $E_{i/f}$ are the eigenvalues of the initial/final orbitals, respectively; $m$ is the electron mass; and $\mathbf{p}$ is a momentum operator. The wavefunctions of the initial/final states $\psi_{i/f}$ are defined as the $\ket{g}$ and $\ket{e}$ states, corresponding to Eq.~\ref{eq:couplingConstant-g}. As the wavefunctions of $\ket{g}$ and $\ket{e}$ states are not necessarily identical, we extract both wavefunctions separately by using the modified version of PyVaspwfc \cite{Davidsson2020}. Furthermore, to differentiate the dipole components, we express $\boldsymbol{\mu}$ as $ \boldsymbol{\mu} = \abs{\mu_x}\hat{x} + \abs{\mu_y}\hat{y} + \abs{\mu_z}\hat{z}. $
This allows us to identify a purely \textit{in-plane} dipole if $\mu_z$ = 0; otherwise, it has an \textit{out-of-plane} component.

\subsubsection{Radiative transition and lifetime}
One of the key performance metrics for a quantum memory is the radiative transition rate, from which a lifetime can be obtained
\begin{equation}
\Gamma_{\mathrm{R}}=\frac{n_D e^2}{3 \pi \epsilon_0 \hbar^4 c^3} E_0^3 \mu^2,
\end{equation}
where $e$ is the electron charge; $\epsilon_0$ is vacuum permittivity; $E_0$ is the energy difference between ground and excited states; $n_{\mathrm{D}}$ is the refractive index of the host material, which is 1.85 for hBN in the visible \cite{10.1021/acsphotonics.8b00127}; and $\mu^2$ is the modulus square of the transition dipole moment obtained by Eq.\ \ref{eq:dipole}. The lifetime $\tau$ is the inverse of the transition rate. It should be noted that this lifetime belongs to the optical transition between $\ket{g}$ and $\ket{e}$ states, which is different from the storage time.

\subsection{Criteria for quantum memory performance} \label{sec:criteria}
Typically, fidelity, memory efficiency, and storage time are essential parameters for assessing the quantum memory performance of any systems. Nonetheless, they are incomplete in the aspect of the prerequisite of electronic structures and compatibility with other systems. Thus, this work has established some further criteria to comprehend all other necessary factors crucial for the performance as follows.
\begin{enumerate}[i)]
    \item Inheriting $\Lambda$ structure: As a quantum memory requires the $\Lambda$ structure, a suitable defect is desired to inherit both triplet and singlet configurations. This way, the triplet configuration allows for the transition between $\ket{g}$ and $\ket{e}$, which is responsible for the signal pulse, while the singlet one allows for the state $\ket{s}$ responsible for the control pulse, which transitions with $\ket{e}$. This triplet-singlet transition exists via the intersystem-crossing channel.
    \item Compatible ZPL with specific application: In principle, each defect can emit a photon with any arbitrary wavelength; hence, we attempted to match its ZPL with other physical quantum systems. This comes from the fact that the quantum memory in NV center struggles with losses for long-distance quantum communication owing to the need for wavelength conversion to couple with photonic devices \cite{PhysRevApplied.9.064031}. Matching the ZPL between hBN defects and other coupling systems can then potentially circumvent this issue. However, we emphasize that the defects with inconsistent ZPL remain meaningful, and of course, their compatibility can still be tuned or even coupled with other systems not stated in this work.
    \item Practical quality factor of a cavity: According to Eq.~\ref{eq:decay-quality}, the required quality factor of the cavity depends on the signal field with optical frequency $\omega$ and coupling constant $g_c$, and the decay rate of a cavity $\kappa$. The quality factor can then lead to high required values. Nevertheless, the physical implementation is limited to only the order of 10$^{3}$ for a dielectric cavity \cite{Vogl2019} and up to 10$^{7}$ for a photonic crystal cavity \cite{pcc}. As a consequence, we aim to identify candidate defects for quantum memory that require a cavity with a quality factor less than 10$^{7}$.
    \item Wide (acceptance) bandwidth: This allows one to relax the requirements of the linewidth of the to-be-stored quantum state, making it easier to mode-match this state and the memory.
\end{enumerate}
It is important to note that fidelity is assumed to be equal to one in this case due to the non-decay and non-dephasing quantum memory model.

\section{Results and discussion}
Among a large number of possible optical transitions, we first indicate the condition from our model for reaching 95\% writing efficiency. Then, the quality factor and bandwidth are calculated for all defects. The defect's electronic transition is in turn revealed by investigating all possible decay paths and collected in the database. The interactive browsing interface for retrieving defect's properties will be accessible in the future.
\begin{figure}[ht]
    \centering
    \includegraphics[width = 0.45\textwidth]{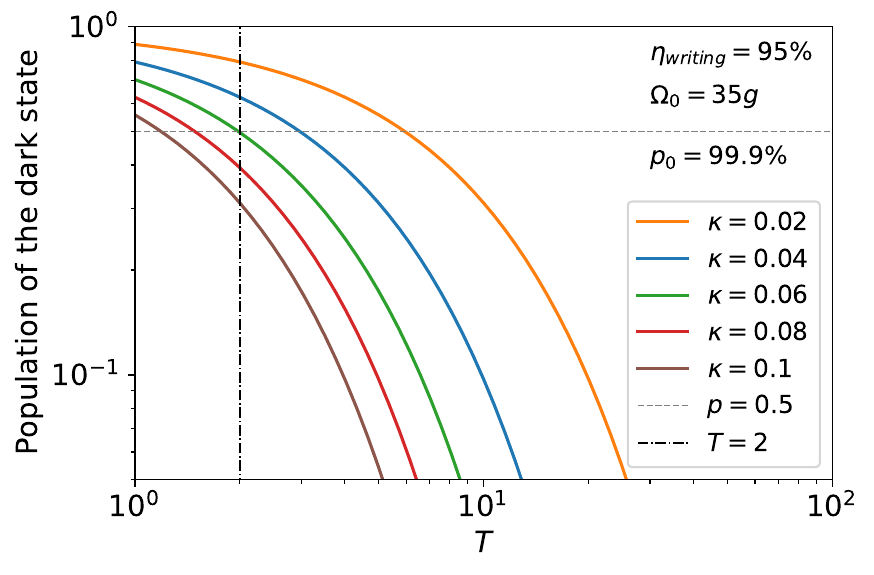}
    \caption{Finding optimal decay rate of a cavity where the initial condition is set to match 95\% writing efficiency and the initial dark state population is 99.9\% initially located at $\ket{g}$. All parameters are scaled with the coupling constant $g_c$.
    }
    \label{fig:decay_rate}
\end{figure}

\begin{figure}[ht]
    \centering
    \includegraphics[width = 0.45\textwidth]{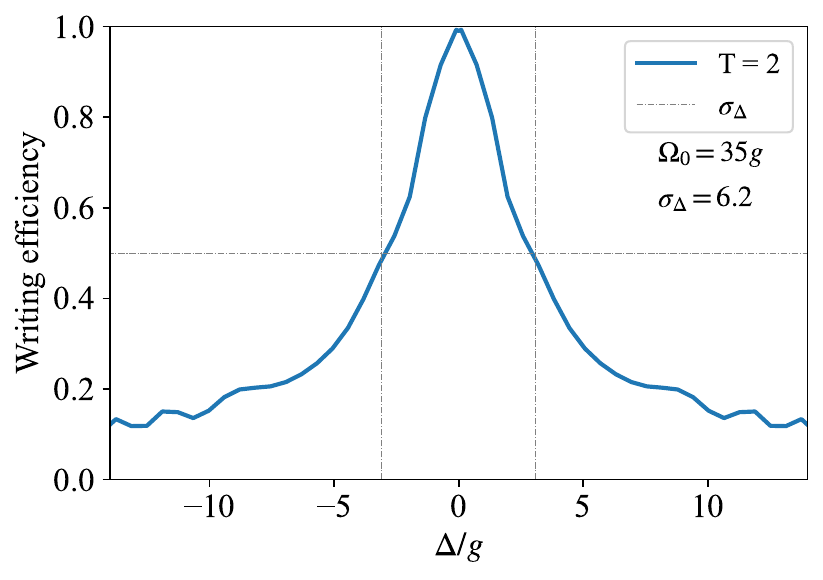}
    \caption{Reduction of writing efficiency affected by one-photon detunning.}
    \label{fig:HWHM}
\end{figure}

\begin{figure*}[ht]
    \centering
    \includegraphics[width = 1\textwidth]{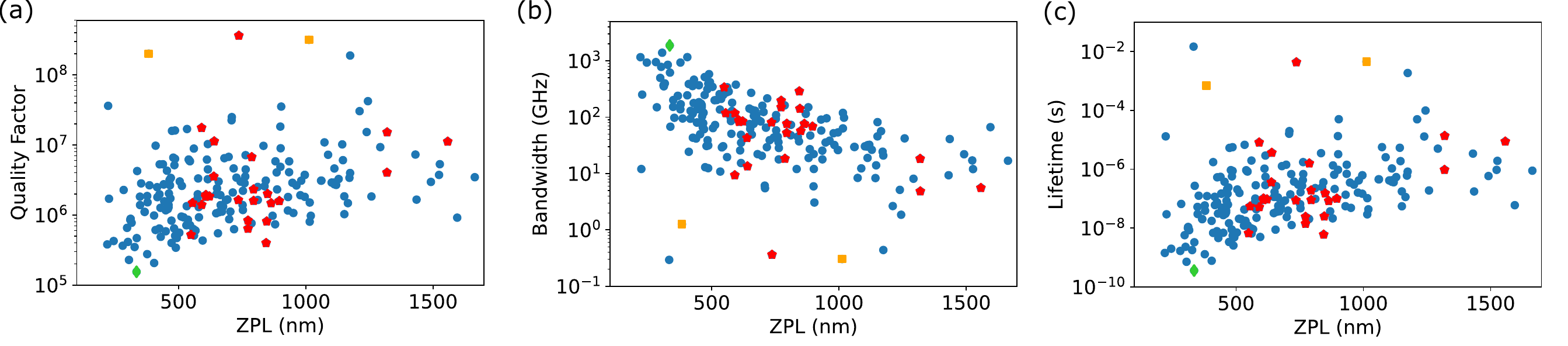}
    \caption{hBN Quantum memory performance based on 257 optical triplet-spin transitions where (a) is quality factor $Q$, (b) bandwidth $\Delta$, and (c) lifetime $\tau$. The green diamond marks a defect with the lowest quality factor and highest bandwidth. The red stars represent the defects compatible with other quantum systems, whereas the orange squares represent the defects with the highest quality factor and lifetime. The remaining blue dots are all other studied defects.}
    \label{fig:defect_performance}
\end{figure*}

\subsection{Condition for writing efficiency} \label{sec:writing_eff}
According to Ref.~\cite{Takla2023}, the condition for a specific writing efficiency was identified. In this work, the condition of $\Omega_0 > 10$ and $T > 2$ for 95\% efficiency is selected for all calculations. This selection of 95\% efficiency can be justified by the fact that the writing efficiency is calculated from the transition probability from $\ket{g}$ to $\ket{s}$ without any decay. Also, the 99.9\% dark state population is assumed to be in $\ket{g}$ initially, and this condition of state preparation is feasible in practice. As a consequence, the writing efficiency is likely to be high. Then, the cavity decay rate was investigated as shown in Fig.~\ref{fig:decay_rate} based on the criterion which the maximum decay rate can remain the dark-state population to be greater than 50\% until the writing process loses the photon at characteristic time $T > 2$. With this condition, it confirms that the quantum memory can achieve 95\% writing efficiency and also beat the no-cloning limit, which is essential for quantum networks \cite{Cho2016,Grosshans2001}.\\
\indent Fig.~\ref{fig:decay_rate} illustrates that the maximum possible decay rate is 0.06$g_c$. Although lower decay rates of the cavity also satisfy the assigned condition, they would yield even higher quality factor, which is unlikely reachable by an ordinary cavity. As a result, we fix the decay rate of 0.06$g_c$ throughout all calculations. Then, by substituting this decay rate, the bandwidth can now be estimated from Eq.~\ref{eq:simplify-bandwidth}. As demonstrated in Fig.~\ref{fig:HWHM}, the bandwidth yields 6.20$g_c$. Then, the properties of each individual defect can be examined and will be discussed in the following section.

\subsection{hBN quantum memory performance}
In this section, we aim to investigate the quality factor and bandwidth among 257 triplet-spin transitions of 211 defects in hBN under 95\% writing efficiency. To map between the quantum memory performance and the material's properties, the coupling constant is computed according to the properties of each defect, consisting of ZPL, transition dipole moment, and radiative transition rate. \\
\indent From the database, we found that if the neutral-charge defect prefers the doublet spin multiplicity, adding or removing an extra electron will rearrange the occupation and then turn the doublet configuration into triplet or singlet ones, respectively. Hence, the database was collected in such a way that if the neutral-charge defect cannot belong to the triplet/singlet spin configurations, an electron will be added or removed. This way we can manipulate the defects to support the $\Lambda$ structure by having triplet/singlet configurations qualified for the quantum memory. This is also plausible in experiment, as can be seen from previous charge-state manipulation mechanisms \cite{Wang2016}. Nonetheless, the charge preference from the defect formation energy is beyond the scope of this work.\\
\indent Fig.~\ref{fig:defect_performance}(a) demonstrates that most hBN defects require the quality factor around 10$^{5}$ $-$ 10$^{8}$ with random ZPL emission. The minimum required quality factor is at 1.55$\times 10^{5}$ by the As$_\text{B}$ defect; however, its ZPL at 335 nm is out of the range of interest for quantum technology applications. We then use the stated criteria in Sec.~\ref{sec:criteria} to narrow down the list of defects to focus only on the ones with ZPLs matching those of other physical systems. The result is listed in Tab.~\ref{table:defect_list}, together with examples of potential applications. Some hBN defects have ZPLs consistent with other single photon sources. Meanwhile, ZPLs of some defects are matched with the storage transition of other quantum memories or wavelengths used for quantum computing, and quantum communication. This suggests the potential comparable integration between hBN-based quantum memory and other components without a need for wavelength conversion. We observed that all defects in Tab.~\ref{table:defect_list} except for Al$_\text{B}$V$_\text{N}^{+1}$ and In$_\text{B}$V$_\text{N}^{+1}$ require a quality factor reachable by a photonic crystal cavity (no more than 10$^{7}$). It should be noted that In$_\text{B}$V$_\text{N}^{+1}$ is ruled out due to the very low radiative transition. This rather suggests that the transition from $\ket{g}$ to $\ket{e}$ in this defect is unlikely, such that this defect has effectively no practical $\Lambda$ structure.
\begin{table*}[ht]
\caption{Properties of 25 selected hBN defects compatible with the emission wavelengths of other quantum systems ($\pm$5 nm) where the quality factor and bandwidth are reported under 95\% writing efficiency condition. The nomenclature of C$_\text{B}$C$_\text{N}$C$_\text{B}$C$_\text{N}$-number is explained in the Supplementary Section S1.}
\begin{tabular}{ccccccccc}
\hline Quantum & Target & Other quantum  & Compatible  & Triplet  & ZPL (nm) & $\tau$ (ns) & Q & $\Delta$ (GHz) \\
application & wavelength (nm) & systems & hBN defects & transition  \\
\hline \multirow{2}{*}{Photon source \cite{PhysRevB.99.075430}} & \multirow{2}{*}{552} &  \multirow{2}{*}{PbV$^-$ (diamond)} & Se$_\text{B}$V$_\text{B}$ &  down & 549.1 & 6.7 & $5.2\times10^{5}$ & 338.9 \\
  & & & Ge$_\text{N}$V$_\text{N}$ &  up & 555.1 & 54.7 & $1.5\times10^{6}$ & 117.9 \\
\hline \multirow{2}{*}{Fraunhofer line \cite{Mostafa-qkd}} &589 & Na-D2  & O$_\text{N}$O$_\text{B}$V$_\text{B}^{-1}$ &  up & 590.8 & 8.2$\times10^3$ & 1.8$\times10^{7}$ & 9.4 \\
  &590 & Na-D1 & S$_\text{B}$V$_\text{B}$ & down & 591.1 & 51.7 & $1.4\times10^{6}$ & 117.6\\
\hline Photon source \cite{GeV} & 602 & GeV$^-$ (diamond) & C$_\text{B}$C$_\text{N}$C$_\text{B}$V$_\text{B}$ & down & 606.7 & 90.3 & $1.8\times10^{6}$ & 87.8 \\
 Memory \cite{Pr-memory} & 606 & Pr$^{3+}$:Y$_2$SiO$_5$ & Ge$_\text{B}$N$_\text{B}$V$_\text{N}$ & up & 607.7 & 101.7 & 1.9$\times10^6$ & 82.7 \\
\hline Photon source \cite{PhysRevLett.119.253601} & 620 &  SnV$^-$ (diamond) & Si$_\text{N}$V$_\text{N}$ & up & 621.6 & 95.0 & $1.9\times10^{5}$ & 84.6 \\
\hline Photon source \cite{PhysRevLett.85.290} & \multirow{2}{*}{637} & \multirow{2}{*}{NV$^-$ (diamond)} & Sb$_\text{B}$ & up & 638.7 & 359.6 & $3.6\times10^{6}$ & 42.9 \\
Memory \cite{NV-memory} & & & C$_\text{B}$V$_\text{N}$V$_\text{B}^{-1}$ & up & 640.3 & 3.6$\times10^{3}$ & $1.1\times10^{7}$ & 13.4 \\
 \hline Computing \cite{Ca-QC} & 729 & Ca$^{+}$ & \multirow{2}{*}{Ga$_\text{N}$} & \multirow{2}{*}{down} & \multirow{2}{*}{735.1} & \multirow{2}{*}{87.8} & \multirow{2}{*}{$1.6\times10^{6}$} & \multirow{2}{*}{80.9}  \\
 Photon source \cite{SiV-SPS}  &\multirow{2}{*}{738} & \multirow{2}{*}{SiV$^-$ (diamond)} & &  &  & &  & \\
 Memory \cite{PhysRevLett.119.223602} & & & Al$_\text{B}$V$_\text{N}^{+1}$ & up & 737.0 & 4.4$\times10^{6}$ & $3.6\times10^{8}$ & 0.4 \\
\hline \multirow{2}{*}{Memory \cite{Rb-memory}} &\multirow{2}{*}{780} & \multirow{2}{*}{Rb-D2} & C-V$_\text{N}$V$_\text{B}$ & up & 773.0 & 14.1 & $6.4 \times10^{5}$ & 196.6 \\
 & & & C$_\text{B}$C$_\text{N}$C$_\text{B}$C$_\text{N}$-3 & down & 773.0 & 24.1 & $8.4\times10^{5}$ & 150.4 \\
\hline Memory \cite{Damon2011} &793 & Tm$^{3+}$:Y$_2$SiO$_5$ & P$_\text{B}$V$_\text{B}^{+1}$ & down & 787.8 & 1.6$\times10^{3}$ & $6.7 \times10^{6}$ & 18.4 \\
\hline \multirow{2}{*}{Memory \cite{Rb-memory}} & \multirow{2}{*}{795} & \multirow{2}{*}{Rb-D1} & C$_\text{N}$V$_\text{N}$ & up & 794.9 & 192.0 & $2.3 \times10^{6}$ & 52.6 \\
 & & & C$_\text{B}$C$_\text{N}$C$_\text{B}$C$_\text{N}$-2 & up & 795.2 & 90.3 & $1.6\times10^{6}$ & 76.7 \\
\hline \multirow{2}{*}{Communication \cite{Jofre2010}} & \multirow{2}{*}{850} &  \multirow{2}{*}{Telecom-1}  & Al$_\text{N}$Al$_\text{N}$ & up & 844.3 & 6.0 & $4.0\times10^{5}$ & 288.5 \\
& & & P$_\text{B}$N$_\text{B}$V$_\text{N}^{-1}$ & down & 846.6 & 25.0 & 8.1$\times10^5$ & 141.3\\
Memory \cite{Cs-memory} &852 &  Cs-D2 & \multirow{2}{*}{C$_\text{B}$C$_\text{N}$C$_\text{B}$V$_\text{N}$V$_\text{B}^{-1}$} & \multirow{2}{*}{up} & \multirow{2}{*}{850.0} & \multirow{2}{*}{152.6} & \multirow{2}{*}{$2.0\times10^{6}$} & \multirow{2}{*}{57.1} \\
Fraunhofer line \cite{Mostafa-qkd} & 854 & Ca-II & & & & & & \\
\hline Memory \cite{SiCmemory} & 862 &  V$_\text{Si}^-$ (silicon carbide) & As$_\text{N}$V$_\text{N}^{+1}$ & up & 864.1 & 83.9 & $1.5 \times10^{6}$ & 76.3 \\
\hline  \multirow{2}{*}{Memory \cite{Cs-memory}} &  \multirow{2}{*}{894} & \multirow{2}{*}{Cs-D1} & In$_\text{B}$V$_\text{N}^{+1}$ & up & 894.4 & 6.2$\times10^9$ & $1.3\times10^{10}$ & 0.0 \\
  & & & C$_\text{B}$C$_\text{N}$C$_\text{B}$C$_\text{N}$-1 & up & 896.0 & 100.8 & $1.6\times10^{6}$ & 68.4 \\
\hline \multirow{2}{*}{Communication \cite{Xu2007}} & \multirow{2}{*}{1330} & \multirow{2}{*}{Telecom O-band} & C$_\text{B}$C$_\text{N}$V$_\text{N}^{-1}$ & up & 1319.8 & 1.4$\times10^4$ & $1.5 \times10^{7}$ & 4.8 \\
&  & & Ga$_\text{N}$N$_\text{B}$V$_\text{N}^{-1}$ & up & 1319.1 & 957.9 & $4.0\times10^{6}$ & 18.3 \\
\hline Communication \cite{Avesani2021} &1550 & Telecom C-band & P$_\text{B}$V$_\text{B}^{-1}$ & up & 1557.8 & 8.8$\times10^3$ & $1.1\times10^{7}$ & 5.6 \\
\hline
\end{tabular}
\label{table:defect_list}
\end{table*}

Further, we compare the performance with the bandwidth as shown in Fig.~\ref{fig:defect_performance}(b). It displays that the quality factor has a reverse relation with the bandwidth. That is, low-quality factor yields high bandwidth. Nevertheless, for all defects in Tab.~\ref{table:defect_list}, except for the defects with unreachable quality factor, their bandwidths are in the order of 10$^{1}$ to 10$^{3}$ GHz. This is already in reach for hBN at room temperature \cite{Vogl2019}.\\
\indent Another essential parameter is the lifetime $\tau$ for each defect as depicted in Fig.~\ref{fig:defect_performance}(c). We found that unusually high lifetime defects (the orange dots) inherit unreachable quality factor by the common cavity and also have narrow bandwidth. This is due to a small transition dipole moment between $\ket{g}$ and $\ket{e}$ states, which results in a low transition rate, as shown in Supplementary Section S2. As a result, these defects are not suitable candidates for quantum memories. Compared to the lifetime of other defects, this suggests that the lifetime likely has a direct correlation with the quality factor, but a reverse correlation with the bandwidth. Considering the defects in Tab.~\ref{table:defect_list}, their lifetime is typically in the order of 10$^{-1}$ $-$ 10$^{-2}$ microsecond, which is sufficiently long compared to the commonly found hBN quantum emitter defects in the order of a few nanosecond.

\begin{figure*}[ht]
    \centering
    \includegraphics[width = 1\textwidth]{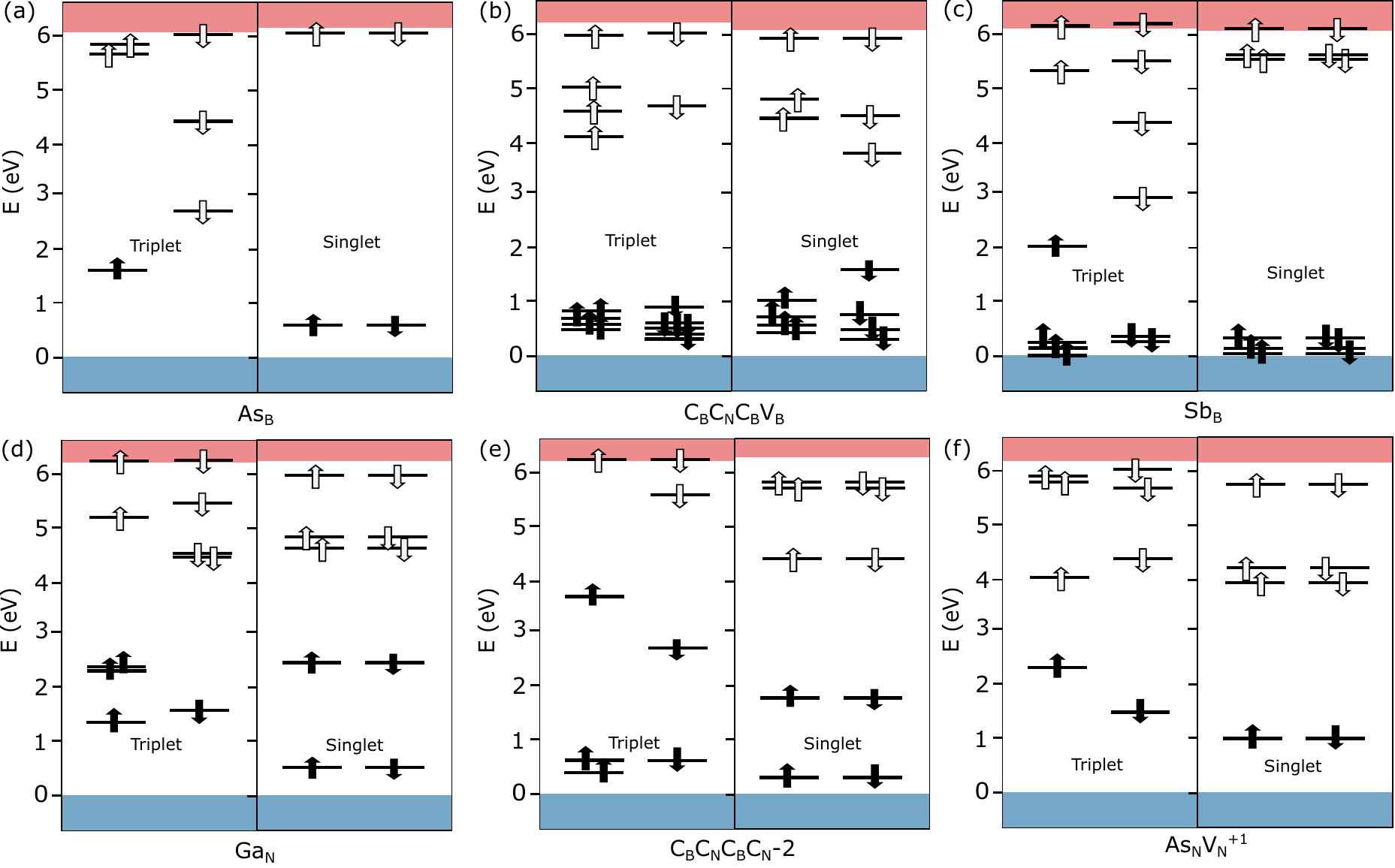}
    \caption{Kohn-Sham electronic transition of triplet and singlet spin configurations where the filled (unfilled) up and down arrows signify the occupied (unoccupied) defect states with spin up and spin down polarization, respectively. (a) As$_\text{B}$, (b) C$_\text{B}$C$_\text{N}$C$_\text{B}$V$_\text{B}$, (c) Sb$_\text{B}$, (d) Ga$_\text{N}$, (e) C$_\text{B}$C$_\text{N}$C$_\text{B}$C$_\text{N}$-2, and (f) As$_\text{N}$V$_\text{N}^{+1}$.}
    \label{fig:band_structure}
\end{figure*}

\subsection{hBN defect properties}
\indent Exemplifying the electronic structures as displayed in Fig.~\ref{fig:band_structure}, we demonstrated both triplet and singlet configurations to ensure that both can be presented. Since the transition of the triplet states can take place via both spin-up and spin-down pathways, as long as the spin-conserved transition is still valid, we split the ZPL calculations into two decay paths: spin-up and spin-down. This is also under the condition that there exist two-level defect states, and the occupied defect state is localized far away from the valence band greater than 1.0 eV. Then, in some triplet-state defects, ZPL energies are reported separately for the spin-up and spin-down transitions.\\
\indent Fig.~\ref{fig:band_structure}(a) shows that As$_\text{B}$ can in principle behave as both triplet and singlet states. However, for the triplet configuration, only the spin-up transition exists and yields ZPL at 335 nm. This is because an occupied defect state of spin-down is absent, which might be difficult for the excitation of a single electron directly from the valence band. This ZPL at 335 nm is relatively low and in turn is impractical for the physical implementation in experiment. For the singlet configuration, the diagram verifies its existence. Likewise, Fig.~\ref{fig:band_structure}(b) shows the electronic transition of C$_\text{B}$C$_\text{N}$C$_\text{B}$V$_\text{B}$. Its triplet-spin electronic states can offer two pathways of transition: spin-up and spin-down. The calculated ZPLs are obtained at 486 nm and 607 nm for spin up and down, respectively (See the Supplementary File for other properties). This implies that only the spin-down transition pathway can give a consistent ZPL with other systems. While the spin-up and spin-down states of its singlet configuration do not align at the same energy level, which is different from As$_\text{B}$, its singlet configuration is still possible. This defect in turn meets the $\Lambda$ structure criterion. As for Sb$_\text{B}$, as indicated in the Tab.~\ref{table:defect_list}, only the spin-up transition has the consistent ZPL with the coupling systems and its singlet configuration is theoretically formable. Thus, this defect is also a promising candidate. For the rest of defects illustrated in Fig.~\ref{fig:band_structure}, they all inherit both triplet and singlet configurations with the consistent ZPL with other systems and are localized well at the mid-gap region. However, the electronic structures of all defects do not have the same characteristics, implying the independence of it on the quality factor and bandwidth. Thus, every single defect needs separate quantum memory performance evaluation.  Note that the coherence time has been considered in this work, as demonstrated in Supplementary Section S4, and yields 0.036 ms. We found that the coherence time is basically independent of the specific defect, which is consistent with previous simulations \cite{Sajid2022,Ye2019}. This comes from the fact that the coherence time $T_2$ is influenced by the magnetic fluctuations of thousands of neighboring nuclear spins such that the ones from the defect do not have a major impact on this. In general, this coherence time can be enhanced by isotopic purification or lattice strain \cite{Ye2019, Lee2022}. Together with the reachable quality factor and wide bandwidths, they all (except for two defects with unreachable quality factor) likely become potential candidates for quantum memory. \\
\indent Having considered the relation between material's properties and quality factor, we found that the ZPL plays a critical role in the resonance frequency $\omega$; however, the more dominant factor to the quality factor of a cavity ($Q$) is the transition rate. As shown earlier, the lifetime ($\tau$) directly correlates with the quality factor. This can be justified by the radiative transition decay rate. That is, the high radiative transition decay rate leads to the high coupling constant ($g_c$) and also high decay rate of a cavity ($\kappa$). This subsequently lowers the quality factor that the cavity needs to achieve. It is also important to note that in principle each defect can have non-radiative decay paths, particularly for the defects having the unpreserved spin polarization for the optical transition between the top-most occupied defect state and the bottom-most unoccupied defect state. Therefore, the non-radiative decay would reduce the radiative decay rate. In this case, a higher quality factor is required (that then enhances the radiative decay via the Purcell effect).\\
\indent Turning to consider the reverse correlation of bandwidths to quality factor and lifetime, it also pinpoints the radiative transition decay rate. That is, the bandwidth depends on the coupling constant between the signal field and the optical transition. According to Eq.~\ref{eq:simplify-bandwidth}, the wider bandwidth is directly proportional to the higher coupling constant. As a result, the high radiative transition rate of a defect can widen the bandwidth.\\
\indent Finally, as the quality factor and bandwidth strongly depend on the transition rate or lifetime of a defect, fine tuning, such as strain engineering, is unlikely to enhance the quantum memory performance since strain predominantly manipulates the ZPL rather than the decay rate. This suggests two possible improvements: (i) applying other defect types instead of fine tuning is a more promising way for a performance improvement, and (ii) fine tuning is important for tailoring the ZPL to couple efficiently with other quantum systems.

\section{Conclusion}
In this work, we investigated the possible implementation of a quantum memory in a variety of hBN defects. By a model to evaluate the performance of the quantum memory, the relation between hBN defect properties and quantum memory performance has been established. Then, DFT calculations were performed to explore a large number of defects and characterized their spin multiplicity as well as ground and excited state properties. The results reveal that the lifetime of defects directly correlates with the quality factor, but reversely correlates with the bandwidth. This suggests that the radiative transition decay rate is a key parameter for quantum memory performance (quality factor and bandwidth). Also, tailoring the performance by fine tuning might be harder than finding optimal defect choices. Among 257 triplet-spin transitions, 25 defects have matched the ZPL of other quantum systems under the $\pm$5 nm variation. Moreover, the majority of hBN defects require an experimentally reachable cavity quality factor at 10$^{5}$ to 10$^{7}$ (for photonic crystal cavities) as well as inherit wide bandwidths for 95\% writing efficiency. This indicates the potential of hBN defect candidates for a quantum memory. In addition, the achieved database can also benefit other quantum technologies, such as ODMR due to the available intersystem-crossing channels between triplet and singlet configurations.\\
\indent Finally, the hBN defects have now been identified to support the $\Lambda$ structure. This becomes qualified for any other mechanism to couple with, such as the nuclear spin. This scenario therefore will be further investigated in details. Our work could therefore provide a promising pathway for relatively simple quantum memories based on room-temperature solid-state photonics. This would be an important milestone due to the role of quantum memories in space \cite{10.1038/s41534-021-00460-9}. The quantum emitter system of hBN is already demonstrated to be compatible with space applications \cite{10.1038/s41467-019-09219-5}. Moreover, our model can be easily used for other color centers in solid-state crystals as well. We expect that with the techniques described in this work, the largely unexplored potential of color centers in materials beyond diamond can be easily evaluated for quantum memories.

\section*{Data availability}
All data from this work is available from the authors upon reasonable request. The DFT database that reveals the triplet and singlet properties is openly available at \href{https://doi.org/10.5281/zenodo.10343468}{https://doi.org/10.5281/zenodo.10343468}.

\section*{Notes}
The authors declare no competing financial interest.

\begin{acknowledgments}
This research is part of the Munich Quantum Valley, which is supported by the Bavarian state government with funds from the Hightech Agenda Bayern Plus. This work was funded by the Deutsche Forschungsgemeinschaft (DFG, German Research Foundation) - Projektnummer 445275953. The authors acknowledge support by the German Space Agency DLR with funds provided by the Federal Ministry for Economic Affairs and Climate Action BMWK under grant number 50WM2165 (QUICK3) and 50RP2200 (QuVeKS). T.V. is funded by the Federal Ministry of Education and Research (BMBF) under grant number 13N16292. C.C. is grateful to the Development and Promotion of Science and Technology Talents Project (DPST) scholarship by the Royal Thai Government. S.S. acknowledges funding support by Mahidol University (Fundamental Fund: FF-093/2567 fiscal year 2024 by National Science Research and Innovation Fund (NSRF)) and from the NSRF via the Program Management Unit for Human Resources \& Institutional Development, Research and Innovation (grant number B05F650024). The computational experiments were supported by resources of the Friedrich Schiller University Jena supported in part by DFG grants INST 275/334-1 FUGG and INST 275/363-1 FUGG. The authors are grateful to Joel Davidsson for the source code of transition dipole moments for two wavefunctions.
\end{acknowledgments}


\bibliography{main}

\clearpage

\end{document}